\documentstyle[11pt]{article}

\newcommand{\et}{$\eta$}
\newcommand{\etb}{$\eta$ }
\newcommand{\myp}{\mbox{$\Delta N \rightarrow N N \eta$ }}

\newcommand{\be}{\begin{equation}}
\newcommand{\ee}{\end{equation}}
\newcommand{\bea}{\begin{eqnarray}}
\newcommand{\eea}{\end{eqnarray}}
\newcommand{\imag}{{\rm Im}}
\newcommand{\cl}{{\cal L}}
\newcommand{\pa}{\partial }
\newcommand{\md}{m_{\Delta}}
\newcommand{\mn}{m_{N}}
\newcommand{\mpi}{m_{\pi}}
\newcommand{\nsb}{$N(1535)$ }
\newcommand{\ns}{$N(1535)$}

\begin{document}
\pagestyle{empty}
\begin{center}
{\Large
PRODUCTION OF ETA-MESONS IN COLLISIONS OF NUCLEONS AND
        DELTA-RESONANCES
        \footnote{\it Work supported by BMFT and GSI Darmstadt}
        \\}
  \bigskip
  \bigskip
  W. PETERS, U. MOSEL, A. ENGEL\\
  {\em Institut f\"ur Theoretische Physik, Universit\"at Giessen\\
       D--35392 Giessen, Germany}\\
\end{center}
\begin{abstract}
We calculate the cross section for the production of $\eta$-mesons via
\mbox{$\Delta N \to N N \eta$} in a relativistic One-Boson-Exchange-Model.
Using this cross section we then determine the probability for the production
of an $\eta$-meson by a $\Delta$-resonance moving in nuclear matter.
The result is compared to prescriptions in BUU-calculations in which
\et-production proceeds both through a direct channel and through the
sequential process $\Delta \to N \pi ; \pi N \to N \eta$.

\end{abstract}
\newpage
\pagestyle{plain}
\section{Introduction}

The detection of mesons produced in heavy-ion-collisions is an important
means for the investigation of the behavior of nuclear matter under extreme
conditions. An analysis of pion spectra, for example, has shown that within
the reaction zone, where at bombarding energies of about 1 GeV per nucleon
baryon densities up to  $3\rho_{0}$ are reached,
about a third of the nucleons  is excited to $\Delta$-resonances
\cite{metag}. Numerical
simulations of heavy-ion-collisions arrive at the same result \cite{moselpb}.

A direct consequence of this resonance matter formation is in these
calculations
an enhancement of sub-threshold particle production \cite{mosel,asbn}.
If the bombarding energy per nucleon is below the free threshold for
the production of a particle,
even when Fermi motion is taken into account, then a large part of  the
observed reaction cross section is, according to numerical simulations,
due to entrance channels involving resonances; these
resonances store energy which can then be used to produce this particle.

While the calculated particle yields generally agree quite well with experiment
when the resonance channels are included, it is clear that this
result depends on the cross sections used for the production processes
involving resonances in the entrance channel; the BUU-calculations mentioned
just use an ad-hoc recipe for the direct reaction channel $\Delta N
\longrightarrow NN \mbox{meson}$ and contain an explicit treatment of the
sequential channel $\Delta \to N \pi ; \pi N \to N \eta$.

It is thus the aim of this study to investigate the elementary production
process in more detail in order to put the results
on the dominance of resonance channels
on a firmer ground. For this study we chose the process \myp
because all the interaction vertices occurring there
are known from other processes.

\section{The process \myp}

We use a relativistic One-Boson-Exchange-Model to calculate the
cross section of the process \myp. Models like this have been used
successfully to describe the production of \et -mesons
in collisions between nucleons [5-7].

\subsection{Couplings}

We use the standard interaction-terms in the hadronic Lagrangian.
Because the $\Delta$ has isospin $3/2$, the only exchange bosons  that have
to be taken into account
are the isovector mesons $\pi$ and $\rho$.

The interaction-terms involving the pion  are
(the subscript $N^*$ always refers
to the $N(1535)$-resonance):
\bea
\cl_{\Delta N\pi} &=&
\frac{g_{\Delta N \pi}}{\mpi}\;\;
\bar{\Psi}_{\Delta}^{\mu} \hat{\vec{T}}  \Psi_{N}\;
\pa_{\mu} \vec{\pi} \;\; +h.c. \nonumber  \\
\cl_{NN^{*}\pi} &=& g_{NN^{*}\pi} \;
                        \bar{\Psi}_{N^{*}}\frac{\vec{\tau}}{2}\Psi_{N}
\; \vec{\pi} \;\; + h.c.
\quad .
\eea
For the interactions containing the $\rho$-meson we use:
\bea
\cl_{\Delta N\rho} &=&
 i \frac{g_{\Delta N \rho}}{\md+\mn}\;\;
\bar{\Psi}_{\Delta}^{\mu}\; \hat{\vec{T}} \; \gamma^{\nu} \; \gamma_{5}\;
\Psi_{N}\;\;
(\pa_{\mu} \vec{\rho}_{\nu} - \pa_{\nu} \vec{\rho}_{\mu})\;\; +h.c.
\nonumber \\
\cl_{N N^{*}\rho} &=&
i g_{NN^{*}\rho} \;\; \bar{\Psi}_{N^{*}} \gamma_{5} \gamma_{\mu}
\frac{\vec{\tau}}{2} \Psi
\;\; \vec{\rho}^{\mu} + h.c.
\quad .
\eea
Finally the coupling of the $\eta$ is taken to be:
\be
\cl_{NN^{*}\eta} = g_{NN^{*}\eta} \;
                        \bar{\Psi}_{N^{*}}\Psi_{N} \;\eta
\;\; + h.c.
\quad .
\ee
The diagrams that are to be considered are shown in Fig. 1 (Top).
We neglect pre-emission graphs
since these
are suppressed kinematically.
Graphs, where the \etb is not produced via a \nsb but via a nucleon,
are neglected as well, because the $NN\eta$-vertex is assumed to
be negligible due to an analysis of Bhalearo and Liu \cite{bhal}.
In principle, there could also be a $N(1535)$-$\Delta$-$\pi$-coupling.
However, according to \cite{pdg} the branching ratio for the decay
$N(1535) \to \Delta \pi$ is smaller than 1\%. Therefore
this coupling is clearly
negligible.

For the off-shell particles in the diagrams in Fig. 1 (Top).
we use monopole form factors:
\be
f(q^2) = \frac{\Lambda^2 - m^2}{\Lambda^2 - q^2}
\ee
\subsection{Parameters}

The value of the $\Delta$-nucleon-pion coupling is easily determined using
the decay-width of the $\Delta$:
\be
g_{\Delta N \pi} = 2.15
\ee
which corresponds to a $\Delta$-width of 115 MeV. This number
is rather fixed because of the well known width of the $\Delta$.

The experimental values of the \ns -width and its branching ratios have
big uncertainties \cite{pdg}; in order to fix
$g_{N N^* \pi}$ and $g_{N N^* \eta}$
we therefore consider the process $\pi N \to N \eta$.
These
coupling-constants were already determined by other groups
(for example \cite{vetter,bhal}). However,
different models or different energy ranges were employed
in these analyses so that we are forced,
in order to
ensure consistency within the model used, to fix these numbers using
our parameterization.

To calculate the cross section for $\pi N \to N \eta$ one has to evaluate
the diagrams shown in Fig. 1 (Bottom). The couplings $g_{N N^* \eta}$
and $g_{N N^* \pi}$ enter in two different ways into these diagrams:
First, they appear
as factors at the vertices and, second, they enter through the \ns -width
in the
propagator. The dependence of the \ns -width and its branchingratio on
these couplings was determined by evaluating the diagrams for the decay into
a pion and an \et, respectively. The two-pion decay of the \nsb was assumed
to give a 10 \% contribution to the total width.

Evaluating the diagrams in Fig. 1 (Bottom) numerically and comparing
these calculations with the data available for $\pi N \to N \eta$
we arrived at the numbers (Fig. 2):
\bea
g_{N N^* \pi} = 1.37  \nonumber \\
g_{N N^* \eta} = 1.53
\eea
The couplings involving the $\rho$-meson are harder to determine.
However, the
following calculations show, that the contribution of
the $\rho$-exchange is much less important than the exchange of a pion.
We, therefore, took the numbers of other groups (e.g. \cite{vetter})
for these couplings:
\bea
g_{\Delta N \rho} = 13.3 \nonumber \\
g_{N N^* \rho} = 0.61
\label{rhoc}
\eea
(see Sec. \ref{dnne}).

\subsection{Kinematics, the exchanged pion}
\label{sabc}

The kinematics of the reaction \myp need further attention, because the
exchanged pion can be on-shell. This can be seen from the diagrams in
Fig. 1 (Top): If the exchanged meson is a pion and one simply cuts
the meson line, one ends up with two diagrams that are both possible
as reactions between on-shell particles. The first one stands for the
$\Delta$-decay and the second one for the process $\pi N \to N \eta$. A
detailed analysis shows, that the pion can be on-shell for cm-energies
of the $\Delta N$-system from about 50 MeV above the threshold
for \et-production
up to about
7 GeV.
In this case the pion propagator becomes infinite,
and the integration over the phase space  of the final particles
cannot be performed because the integral diverges.

The solution to this problem is to remember that the reaction
considered here
does not take place in the vacuum, but during a heavy-ion-collision, i. e.
within a medium with a non-vanishing baryon density. In such a medium the
pion acquires an in-medium width which renders the pion propagator finite, so
that the total cross section can be calculated.

This width enters into the propagator via the pion selfenergy $\Sigma_{\pi}$:
\be
G_{\pi} = \frac{i}{q^2 - m_{\pi}^2 - \Sigma_{\pi}}
\quad .
\ee
The relation between the width and the imaginary part of the pion selfenergy
is:

\be
\imag \;\Sigma_{\pi} = - E_{\pi}\; \Gamma_{\pi}
\label{impi}
\quad .
\ee
The width of the pion is nothing else but its inverse lifetime and can
therefore be expressed by means of the mean free path:
\be
\Gamma_{\pi} \; =\; \frac{1}{\tau} \; =\; \frac{v_{\pi}}{l}
\; =\; v_{\pi} \; \rho \; \sigma_{\pi N} + \tilde{\Gamma}_\pi  (\rho)
\quad ,
\label{gampar}
\ee
where $v_{\pi}$ is the pion velocity, $\rho$ is the nuclear density and
$\sigma_{\pi N}$ is the total pion-nucleon cross section, which
contains elastic scattering as well as inelastic processes \cite{lb}. By
using the form (\ref{gampar}) we assume a local-density approximation for
the selfenergy of the pion; we are aware of the fact, that this is a good
approximation only in the nuclear bulk matter region.

$\tilde{\Gamma}_\pi  (\rho)$ in Eq. (\ref{gampar}) represents the
part of the
selfenergy  of the pion which is due to  many-body effects in the nuclear
medium. The effect of 2- and 3-body absorption on the pion selfenergy in the
region of the $\Delta$-resonance has been discussed extensively in the
literature (e.g. \cite{oset,salcedo}). In the present case however
the pion-nucleon
system has an invariant mass above the \et-threshold where several resonances
contribute (N(1440), N(1520), N(1535),\dots),
so that there is no
quantitative information available.
In the absence of any reliable information on the
many-body part of the pion selfenergy we
have therefore simply multiplied the $\pi N$ cross section with a
density dependent factor by making the replacement
\be
\sigma_{\pi N} \longrightarrow \sigma_{\pi N}(1+\alpha(\rho))
\label{alpha} ~.
\ee
with $\alpha(\rho=0) = 0$.
Since we will work at constant density (local density approximation),
effectively this just amounts to multiplying the cross section
$\sigma_{\pi N}$ with a constant factor.

The expression for the imaginary part of the selfenergy
in Eq. (\ref{impi})
can be derived by  evaluating the lowest order diagrams contributing
to $\Sigma_{\pi}$ by means of Cutkosky's rule.
For $\alpha = 0$ it reduces to the well known
form $\imag \;\Sigma_{\pi} = - m_{\pi}\; \Gamma_{\pi}$ if one
chooses the rest frame of the pion as the
reference frame. In our case the more general form (\ref{impi})
must be used, because the width given by (\ref{gampar}) uses the
rest frame of the medium as  the reference frame.

For the real part of the selfenergy we took results of calculations
employing the $\Delta$-hole-model \cite{ewe}.
The $\Delta$-hole-model also yields an imaginary part of $\Sigma$
which is similar to (\ref{gampar}) \cite{oset}.
However, $\sigma_{\pi N}$ then
only contains the $\Delta$ and no other resonances. We improved this
by using a
parameterization of  the experimental data for $\sigma_{\pi N}$ in
(\ref{gampar}), thereby
including also higher resonances.

An important consequence of Eq. (\ref{gampar}) is that because of the density
dependence
of the pion-width the cross section for \myp  depends on the density
as well.
This will be discussed further in the next section.

\subsection{Results}
\label{dnne}

The results of our calculations for the total cross section are given in
Fig. 3 (Top) for three different nuclear densities and $\alpha$ of
Eq. (\ref{alpha}) set to zero.
Interpreting the
density dependence of the cross section at fixed $\sqrt{s}$ one has to
bear in
mind, that a cross section in general is a transition amplitude normalized
with respect to the flux of the incoming particles and the number of
target particles. So the cross section for $\Delta^- p \to n n \eta $ as
plotted in Fig. 3
does not contain the probability for a $\Delta$ meeting a nucleon, which
does, of course, also depend on the density.

Keeping this in mind the density dependence of the cross section at fixed
$\sqrt{s}$ is easy to understand.
It becomes smaller with
increasing $\rho$, because the probability for the exchanged pion to
react with the medium grows with the density.
In the $\sqrt{s}$-region where the pion can be on-shell (see Sec.
\ref{sabc})
the total cross section scales roughly with $1/ \rho$ and diverges for
vanishing density.

Fig. 3 (Bottom) shows the cross section with and without
the exchange of a $\rho$-meson.
One sees that the process \myp is clearly
dominated by the exchange of the pion (The coupling constants used
were given in  Eq. (\ref{rhoc})).
A more recent analysis \cite{schaefer} finds a $\Delta$-$N$-$\rho$-coupling
which is
about a factor of two below the one given in Eq. (\ref{rhoc}), so that
the contribution of the $\rho$-exchange is even smaller.
The large contribution of the $\rho$-exchange found for $NN\to NN\eta$
in \cite{laget,wilkin} does not appear here, because the on-shell pion
dominates the reaction.

\section{Comparison to the BUU-Parameterization}

The purpose of this work is to provide microscopically justified input
for transport theories like the BUU-model to describe
the production of \et -mesons in heavy ion collisions. In standard
BUU-calculations \cite{wolf} the process \myp is implemented in two additive
ways:
\begin{description}
\item[1. Sequential production:]A $\Delta$-resonance can decay into a
nucleon and
a pion, that is treated as a real, on-shell particle. One of the possible
reactions for this pion is then $\pi N \to N \eta$.
\item[2. Direct production:]One of the reactions allowed for
the $\Delta$ besides the decay is the direct production of an \et -meson
via \myp. The cross section for this process was assumed to be equal to
the cross section for $NN \to N N \eta$
at the same $\sqrt{s}$, corrected for the different spin-isospin degeneracies
in the entrance channel.
\end{description}

In the last section it was shown, that the total cross section is
governed by the exchange of an in-medium pion with a finite width.
Consequently the contribution to the \et -production by on-
and off-shell pions cannot strictly be separated. Indeed, Salcedo et al.
\cite{salcedo} have shown for a similar problem
that in the limit of vanishing density the
sequential and the direct production processes give exactly the same result.
Thus, a BUU-calculation employing both the sequential and the direct
production channel might contain some double counting. This is the point
we want to investigate in the following discussions.

In order to do so we consider a single $\Delta$-resonance of given energy
moving in
homogeneous matter with density $\rho$ and calculate the $\eta$ production
rate by solving appropriate rate equations. For simplicity we
take one definite isospin channel, namely a $\Delta^-$ moving within
a medium of protons. In this channel the additional
cross sections we will need below are experimentally well known. Also
for simplicity we first set $\tilde{\Gamma}_\pi  (\rho)$ of
Eq. (\ref{gampar})
equal to zero, thus neglecting many-body reactions of the pion;
these effects will be discussed  separately at
the end of this section.

The rate equations have to contain the decay widths and reaction rates
of all  possible processes. These are for the
\begin{description}
\item[$\Delta$ channels:]

\begin{description}
\item[]
\item[$\Delta$-decay:] This enters into the rate equations via the decay width
${\Gamma\!}_{\Delta}$. In  our scenario there is no medium modification of
this width due to  Pauli blocking, because a $\Delta^-$ can only decay
into a $\pi^-$ and a neutron, for which there is no Pauli blocking
in a medium of protons.
\item[$\Delta$-absorption:] Given the cross section for $\Delta N \to NN$
\cite{wolf} the reaction rate for this reads:
\be
{\Gamma\!}_{\Delta p\to nn} = v_{\Delta}\;\sigma_{\Delta p \to nn}\; \rho
\ee
where $v_{\Delta}$ is the velocity of the $\Delta$ and $\rho$ is the
density of the protons.
\item[Direct \et -production:]The reaction rate for this is given by:
\be
{\Gamma\!}_{\Delta p\to nn \eta} = v_{\Delta}\;\sigma_{\Delta p \to nn\eta}
                                        \; \rho ~.
\label{delrate}
\ee
\end{description}
\end{description}

So far, no  sequential production is included. To account for it
we have to consider the reactions that the pion produced by the decay
of the $\Delta$ can undergo:

\begin{description}
\item[Secondary $\pi$ interactions:]

\begin{description}
\item[]
\item[Reaction with the medium:]Using the total pion-nucleon cross
section, which appeared already in Eq. (\ref{gampar}),
the reaction rate for this is:
\be
{\Gamma\!}_{\pi p} = v_{\pi}\;\sigma_{\pi p}\; \rho
\label{gampin}
\ee
As has been said before, being the total cross section $\sigma_{\pi p}$
contains elastic and inelastic processes.

\item[\et -production:]
Our calculation for the cross section for $\pi p \to n \eta$  are shown in
Fig. 2 together with experimental data.
With that cross section the reaction rate is:
\be
{\Gamma\!}_{\pi p \to n \eta} = v_{\pi}\;\sigma_{\pi p \to n\eta}\; \rho
\ee
\end{description}
\end{description}

Combining all these channels we arrive at the following rate equations:
\bea
\frac{d}{dt} N_{\Delta}(t)&=& \left( -{\Gamma\!}_{\Delta p\to nn}
               -{\Gamma\!}_{\Delta} - {\Gamma\!}_{\Delta p\to nn \eta}
                \right) \; N_{\Delta}(t) \nonumber \\
\frac{d}{dt} N_{\pi}(t)&=& {\Gamma\!}_{\Delta}\; N_{\Delta}(t)
               -{\Gamma\!}_{\pi p}\; N_{\pi}(t)
               \nonumber \\
\frac{d}{dt} N_{\eta}(t)&=& {\Gamma\!}_{\Delta p\to nn \eta }\; N_{\Delta}(t)
               +{\Gamma\!}_{\pi p\to n \eta}\; N_{\pi}(t)
\quad .
\label{rateeq}
\eea
These equations constitute a first collision model; as such they do not
contain any refeeding of the $\Delta$ or of the $\pi$. This is not only
sufficient for our
purpose of comparing the BUU-prescription with our microscopic calculation,
but is even a cleaner test because it does not sum over complicated collision
histories with their inherent averaging over many processes and cross sections.

The first of Eqs.\ (\ref{rateeq}) contains some double counting,
because on-shell pions appear both in
$\Gamma_\Delta$ and $\Gamma_{\Delta p \rightarrow n n \eta}$.
However,
for normal nuclear density it is
$\Gamma_{\Delta p \rightarrow n n \eta} \approx 0.01 \:\:\Gamma_\Delta$,
so that this double counting has negligible
consequences.

Because of the absence of any refeeding terms in the $\Delta$ and $\pi$
channels Eqs. (\ref{rateeq}) can easily be solved.
If one uses the initial conditions
$N_{\Delta}(0)= 1  ; N_{\pi}(0)= N_{\eta}(0)= 0$ and  takes the limit
$P_{\eta} = \lim_{t\to \infty} N_{\eta}(t)$, then $P_{\eta}$ is the
probability for the production of an \et -meson in the first reaction of
either the $\Delta$ or the pion produced in the decay of that $\Delta$
with a proton of the medium.

If the density dependent cross section calculated above is used
in Eq. (\ref{delrate})
for
${\Gamma\!}_{\Delta p\to nn \eta }$, the width
${\Gamma\!}_{\pi p \to n \eta }$
has to be set equal to zero, because then the cross section
$\sigma_{\Delta p \to nn\eta}$ already contains all kinematically possible
invariant masses of the intermediate pion. The results of this calculation
are shown in Fig. 4 (upper curves).
The density dependence of $P_{\eta}$ is weak because the approximate
$1/\rho$-dependence of the cross section is multiplied with the
$\rho$-dependence of the collision rate (Eq. \ref{delrate}).

Also shown in Fig. 4 is the result for the sequential
process, obtained by setting  $\sigma_{\Delta p \to nn\eta}$ to zero in the
rate equations given above (lower curves).
For $\rho = \rho_0$ the sequential \et-production
probability is about 30 \% smaller than the direct production cross section
given by the upper curves. For decreasing density, however, the upper and
the lower curves converge and for $\rho \to 0$ they coincide, in agreement
with the theoretical results of Salcedo et al \cite{salcedo}.
The physical reason for this
is that for $\rho\to 0$ the pion has to travel an arbitrarily long distance
before
reacting with a nucleon. This can only happen, if the pion is strictly
on shell, so that the sequential description becomes exact, and the upper
and lower curves in Fig. 4 have to coincide. The difference between
the two sets of curves thus reflects an important in-medium effect.

It can also be seen from Fig. 4, that at finite density the direct
$\eta$-production and  the sequential process
have different thresholds.
This is due to the fact that the sequential threshold lies above the absolute
energy threshold for $\eta$ production in the $\Delta N$ channel because
the requirement of an on-shell pion constitutes an additional constraint on
the reaction dynamics (see Sec. \ref{sabc}). Only for $\rho=0$ both
thresholds agree, in agreement
with the theoretical considerations of Salcedo et al.

The Fermi momentum of the protons is not taken into account in
Fig. 4. Including this leads to a smearing of all curves over
a certain energy range, which blurs the differences shown
in Fig. 4.

In order to check the validity of the parameterization
of the direct production cross section and the inclusion of
both direct an sequential production channels
in standard
BUU-calculations we now compare our results with
calculations that treat the possible reaction channels like
the BUU-model. In the context of our scenario this means that we have
to include in the rate equations both the sequential production as in Fig. 4
and the direct production process, but now with the density-independent
parameterization given in \cite{wolf} for $\sigma_{\Delta p \to nn\eta}$.
This does not change the result in the limit of $\rho \to 0$ because
in this case ${\Gamma\!}_{\Delta p\to nn \eta }$ vanishes
while ${\Gamma\!}_{\Delta}$ remains constant (first of Eqs. \ref{rateeq}).
The two contributions together with the resulting
total production probability are shown in Fig. 5.

Fig. 6 (Top) shows a comparison of the results of the BUU-simulation
including both the direct BUU-type production and the sequential channel
(i.e. of the results of Fig. 5) with the results of the calculations
employing our density dependent cross section for the direct channel only.
The two curves are remarkably similar although there is some difference
in their energy dependence.
This difference is mainly due to a somewhat different energy dependence
of the calculated cross section for \myp and its BUU-parameterization.
Taking the Fermi motion into account diminishes this difference significantly
(Fig. 6 (Bottom)). The agreement between the two prescriptions at
other densities than
$\rho_{0}$ is similar.

So far, all the calculations have neglected the contribution of
many-body effects on the pion selfenergy. Fig. 7 now shows results
for two different values of $\alpha(\rho_{0})$ of Eq. (\ref{alpha}),
entering via the pion selfenergy and Eq. (\ref{gampin}), respectively.
In both cases the
agreement between the BUU prescription and the calculation with our
calculated cross section is still very reasonable for low to
intermediate kinetic energies of the $\Delta$ ($T_{\Delta} < 1$ GeV). This
agreement
for different values of $\alpha$ is due to the fact that the replacement
(\ref{alpha}) affects the results of both models in an analogous way
by causing a higher reaction probability of the pion.
Only at higher energies we obtain a stronger dependence of
$P_{\eta}$ on $\alpha(\rho_{0})$ than in the BUU-case. This is simply due to
the fact that in the BUU-case the cross section for the direct process
is taken to be density-independent,
so that in this case only the sequential production depends on $\alpha$.

\section{Conclusion}

We have calculated the cross section for \myp in a microscopical model
using empirical interaction vertices.
Because of the special kinematics of this reaction, the exchanged pion
can be on-shell so that a straightforward evaluation of the corresponding
Feynman graphs leads to divergent results.
By including in-medium effects, however, we have obtained a finite,
density dependent result.
This result contains all kinematically possible situations, including
the point, where the invariant mass of the exchanged pion is equal to
its free mass.

We have then used this cross section to check the ad-hoc treatment of the
$\Delta N$ channel for particle production in transport theories like the
BUU-model.
This is essential, because at energies far below threshold $\eta$-mesons
are created mainly via the resonance channel.

The BUU-prescription allows both the direct and the sequential process.
The sequential production involves a pion which is propagated as an on-shell
particle, while the direct process is included on an ad-hoc basis with a
guessed cross section to account for the $\eta$-production via off-shell
pions. In order to compare our `exact` calculation with this BUU-recipe
we have set up a
simple scenario in which the probability for the initial production of an
$\eta$-meson by a $\Delta$ of given energy was calculated for both
cases by solving rate equations.

As a result we have found, that by  allowing sequential production alone
one obtains an  $\eta$-yield, which was clearly below the
one obtained with the cross section calculated in this work (Fig. 4).
Only if we included sequential
and direct production at the same time, using the cross sections as given
in \cite{wolf}, the correct $\eta$-production probability was
reproduced quite well (Fig. 6, Top). The different energy dependence still
visible there vanished nearly completely when the Fermi distribution
of the medium was taken into account (Fig. 6, Bottom).
We thus conclude, that there is no double counting in the
BUU-treatment and that the BUU-recipe for the $\Delta N \rightarrow NN \eta$
cross section works surprisingly well.
This result puts
the enhancement of far-subthreshold particle production obtained in
BUU-calculations on a firmer basis.

\newpage
\begin{figure}
\label{dnnegr}
\caption{Top: Diagrams dominating the process \myp.
         Bottom: Diagrams contributing to the  process $\pi N \to N \eta$.  }
\end{figure}
\begin{figure}
\label{sgpnne}
\caption{Experimental data for the
cross section for  $\pi^- p \to n \eta$ together with our
calculations like described in the text. The data were taken from
\protect{\cite{lb}}.}
\end{figure}
\begin{figure}
\label{sgdnne}
\caption{Total cross section for $\Delta^- p \to n n \eta$.
Top: Assuming three different nuclear densities.
Bottom: With and without the $\rho$-exchange ($\rho = \rho_{0}$).
}
\end{figure}
\begin{figure}
\label{cofm}
\caption{\et -production probability for different densities using only
the direct process with the density dependent cross section (upper curves)
and using only the sequential production (lower curves) without Fermi motion.}
\end{figure}
\begin{figure}
\caption{
Total \et -production probability in a BUU scenario (solid curve).
The lowest (dotted) curve gives the direct production probability obtained
using the BUU-parameterization. The dashed curve shows the sequential
contribution (same as in Fig. 4).
}
\end{figure}
\begin{figure}
\label{cparofm}
\caption{Top: \et -production probability  using only
the direct process with the density dependent cross section calculated in
this paper and using the sum of the sequential and the direct process in the
BUU-parameterization (for $\rho = \rho_{0}$).
Top:  Without Fermi motion.
Bottom: With Fermi motion.
}
\end{figure}
\begin{figure}
\caption{\et -production probability
for two different values of $\alpha(\rho_{0})$ of Eq.
(\protect{\ref{alpha}})
(with Fermi motion, $\rho = \rho_{0}$) }.
\end{figure}
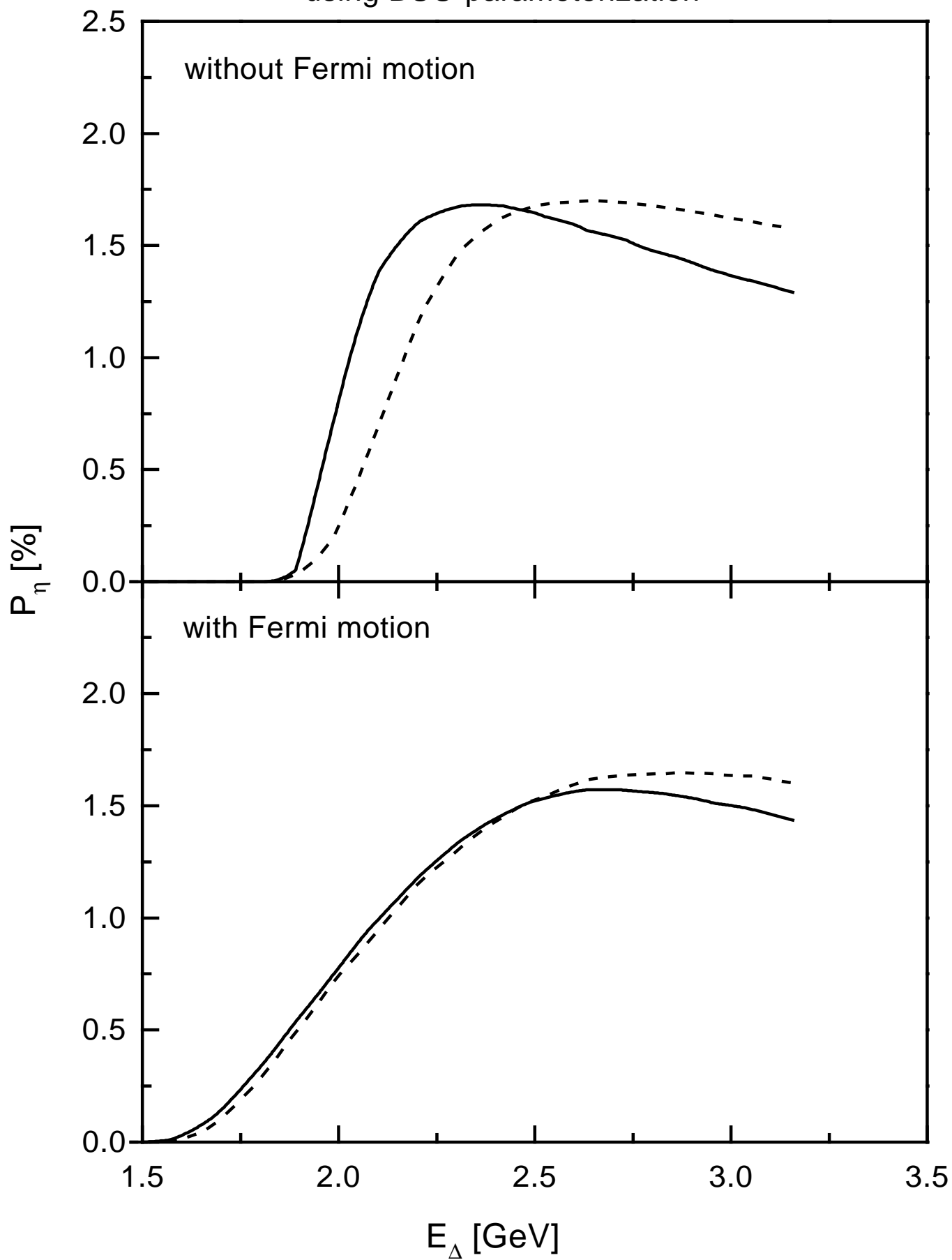
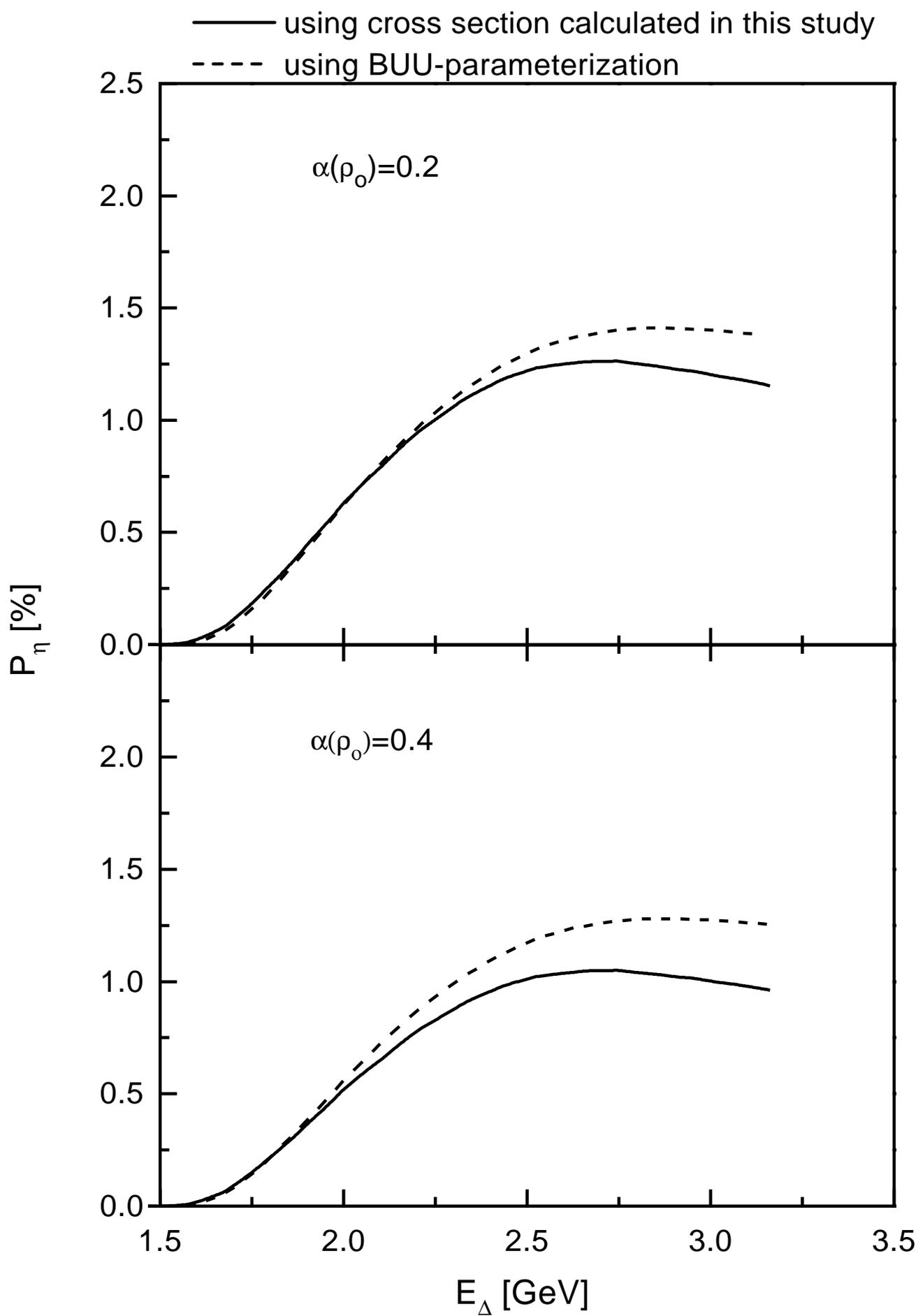
\end{document}